\documentclass{aa}
\usepackage{graphicx}

\begin{document}

\title{Photometric variability in the old open cluster M\,67
\thanks{Tables 1 and 3 are only available in electronic form via
anonymous ftp to cdsarc.u-strasbg.fr (130.79.125.5) or via
http://www.edpsciences.org}}
\subtitle{II.\ General survey\thanks{WIYN Open Cluster Study.\ IX.} }

\authorrunning{Stassun et al.}
\titlerunning{Photometric variability in M\,67}

\author{Keivan G.\ Stassun\thanks{Hubble Fellow}\inst{1}
\and 
Maureen van den Berg\inst{2}
\and
Robert D.\ Mathieu\inst{1}
\and
Frank Verbunt\inst{3} }

\institute{Department of Astronomy, University of Wisconsin, 475 N Charter St, Madison WI 53706, USA
\and
Osservatorio Astronomico di Brera, Via E. Bianchi 46, 23807 Merate (LC), Italy
\and
Astronomical Institute, P.O. Box 80000, 3508 TA Utrecht, The Netherlands}

\offprints{Keivan Stassun, \email{keivan@astro.wisc.edu} }

\date{Received date / Accepted date}

\abstract{We use differential
CCD photometry to search for variability in $BVI$ among 990 stars
projected in and around the old open cluster M\,67. In a previous paper
we reported results for 22 cluster members
that are optical counterparts to X-ray sources; this study focuses on
the other stars in our observations. A variety of sampling rates were
employed, allowing variability on time scales ranging from $\sim 0.3$
hours to $\sim 20$ days to be studied.  Among the brightest sources
studied, detection of variability as small as $\sigma\approx$ 10 mmag is
achieved (with $>3\sigma$ confidence); for the typical star observed, 
sensitivity to variability at levels 
$\sigma\approx 20$ mmag is achieved.  The study is unbiased for 
stars with $12.5<B<18.5$, $12.5<V<18.5$, and $12<I<18$ within a radius 
of about 10 arcmin from the cluster centre. In addition, stars with
$10<BVI<12.5$ were monitored in a few small regions in the cluster.
We present photometry for all 990 sources studied, and
report the variability characteristics of those stars found to be
variable at a statistically significant level. Among the variables, we
highlight several sources that merit future study, including stars
located on the cluster binary sequence, stars on the giant branch, blue 
stragglers, and a newly discovered W\,UMa system.
}

\maketitle

\section{Introduction}
Photometric variability serves as an important means of
identifying stars 
of interest to the study of stellar structure and evolution. 
Origins of stellar photometric variability include 
dynamo-generated magnetic activity, stellar pulsation, and
tidal effects related to stellar multiplicity. 

With the intent of studying the photometric variability of optical
counterparts to known X-ray sources scattered throughout the old
open cluster M\,67, we have obtained sensitive photometry
of stars projected in a roughly square region one-third 
degree on a side centered $5'$ north of
the cluster centre. The results of these observations for the
X-ray sources are described in Paper~I of this series (\cite{vdbergea2001b}).
%
Light curves for nearly 1000 other stars were produced in the
course of our analyses.  In this paper we present the results of an
extensive time-series analysis of the 968 stars included in our
observations that are not optical counterparts of X-ray sources
known to be members of the cluster.

Our basic goal is to identify those stars that exhibit statistically
significant photometric variability of any kind. At the old age of
M\,67 (4 Gyr; \cite{pols}), most single stars rotate too
slowly to exhibit strong dynamo-generated activity, and the
sensitivity of our photometry is insufficient to detect the extremely
low-level ($\sim$ few $\mu$mag) variations that may arise from
solar-analog {\it p}-mode oscillations (\cite{woodhuds}).
Thus, photometric variability in our observations may be an indicator
of, e.g., binary interaction (eclipses or ellipsoidal variations), 
spot-modulated stellar rotation, or stellar activity at levels
not detectable in existing X-ray surveys. Periodic variability is
especially interesting in these contexts, as a periodicity in the
light curve may be fundamentally related to a stellar rotation period,
a binary orbital period, etc. But even if a periodicity is not
apparent in our data, the detection of photometric variability may
point the way to objects that merit further
study.  


Basic photometry has been performed in M\,67 by several authors.
\cite{montea} have presented a deep
($V \sim 20$) colour-magnitude diagram of the central one-half degree
of the cluster, and \cite{fan} have presented spectrophotometry
of similar depth
from 0.4 $\mu$m to about 1 $\mu$m for stars in a $2^\circ \times 2^\circ$
region centered on the cluster. In addition, time-series analysis has 
been performed in M\,67, most notably by
\cite{gillea}, who conducted a very sensitive ($\sim 100
\mu$mag), highly temporally sampled ($\sim 1$ min$^{-1}$) study of
stars in the core of the cluster, resulting in several 
detections of W\,UMa systems and $\delta$ Scuti variables,
as well as tentative detections of other stellar oscillations.
This study was confined for the most part to the central few
arcmin of the cluster. 

The present study complements and extends these previous studies by
combining a study of variability (sensitive to $\sigma\sim$ 10--20 mmag) 
with a reasonably deep ($V \sim 18.5$) colour-magnitude diagram for stars
covering a large area around the cluster centre. Furthermore, the
ongoing CfA spectroscopic survey 
(e.g.\ \cite{lathmathea}) has identified numerous
spectroscopic binaries in the cluster among stars brighter than about
$V\sim 14$; in this study we incorporate the available knowledge of
binarity into our analyses where appropriate.

In Sect.~\ref{m67phot2:data}, we summarise the photometric data and
their analyses. We provide photometry for all of the
sources included in our observations, as well as cross-identifications
of our sources with those of several previous authors. In
Sect.~\ref{m67phot2:results}, we present the basic results of this
study, including an identification of stars exhibiting statistically
significant photometric variability, and a colour-magnitude diagram of
all sources studied. 
The colour-magnitude diagram we present shows a well-defined main
sequence extending to the limit of our photometry, and an evident
binary sequence. Many of the stars identified as photometric variables
lie either on the binary sequence, in the region of the blue stragglers,
and on the giant branch. 
Still other stars lie on the cluster main sequence and exhibit
apparently irregular photometric variations.
Photometric variability at the
levels to which this study is sensitive ($\sim$ 1--2\%) is uncommon
in this cluster among stars that are not known to be X-ray sources or 
binaries (variability having an occurrence rate of only $\sim$ 3\% 
among the non--X-ray sources and non-binaries observed by us). 
We discuss the results of select individual
sources in greater detail in Sect.~\ref{m67phot2:disc}, and summarise
our findings in Sect.~\ref{m67phot2:summ}.

\section{Data and analysis} \label{m67phot2:data}
In this section we provide a summary of our observations and of the
procedures used in their analysis.  For complete details of the data
and of the analysis procedures, the reader is referred to Paper~I.

\subsection{Observations}
Differential $B$$V$$I$ photometry of M\,67 was performed during five
separate epochs with a total time span of two years. The observations
were obtained with 1-meter telescopes at three different observing
sites, each with a different field of view (ranging from $3\farcm8$ to 
$23'$ on a side)
and under a range of observing conditions. The five observing runs
differ considerably in time span and sampling frequency; the shortest
run is the most highly sampled, spanning 2 days with individual
measurements taken at roughly 5-minute intervals, while the longest
run spans nearly 25 days with measurements taken at intervals of
several hours. As each observing run had as its primary target a
different set of X-ray sources, the five runs differ in depth and
range of stellar magnitudes covered.
We encourage the reader to consult the map (Fig.~1) and table
(Table~1) in Paper~I that more fully describe these observation details.

The result is a database of differential photometric measurements for
990 stars in a region roughly $23'$ on a side, centred
approximately $5'$ north of the cluster centre. The database is
complete in this large area for stars with $12.5<B<18.5$,
$12.5<V<18.5$, and $12<I<18$.  In addition, the database includes
stars with $10<BVI<12.5$ in a few small regions within this larger
area.  Some $U$-band photometry was obtained as well, but as only a
small number of stars in a few select regions were observed we do not
include the analysis of the $U$-band data here.

Due to the differences in depth and areal coverage of the five
observing runs, there is relatively little overlap of stars among the
five epochs of data. Thus the light curve of a given star 
typically spans
anywhere from 2 to 25 days with the time-sampling depending
on the particular epoch contributing data to that star. Furthermore,
stars near the bright or faint extremes of the database may not be
present in all three filters depending on the stellar colours.

\subsection{Data reduction and light curve solution}
The roughly 1200 data frames produced in the course of the five
observing runs were reduced using standard IRAF\footnote{IRAF is 
distributed by the National Optical Astronomy Observatories,
which are operated by the Association of Universities for Research
in Astronomy, Inc., under cooperative agreement with the National
Science Foundation.} procedures.  All
stellar sources with S/N $>10$ were identified with
the {\sc daophot} task and aperture photometry was performed using the
{\sc apphot.phot} task.  For the purpose of cross-identifying stars
in our frames with previous work, astrometry was extracted for all stellar
sources identified using the {\sc stsdas.gasp} package, producing
astrometric solutions with formal internal uncertainties of
approximately $0\farcs1$ in each direction. Astrometric positions are
tied to the coordinate system adopted by \cite{montea},
resulting in an external uncertainty of approximately $0\farcs4$
in each direction.

In Table~\ref{tab1} we present the master
list of 977 stellar sources included in our observations, sorted
in order of increasing right ascension, which we were able to 
cross-identify with the database of \cite{fan}. The primary stellar 
identification is that of \cite{fan}, and cross-identifications
with the studies of various other authors are also given. Various
cluster membership estimates from those studies are also provided, when
available.

\setcounter{table}{0}
\begin{table*}[ht]
\centerline{\includegraphics[bb=35 269 610 520]{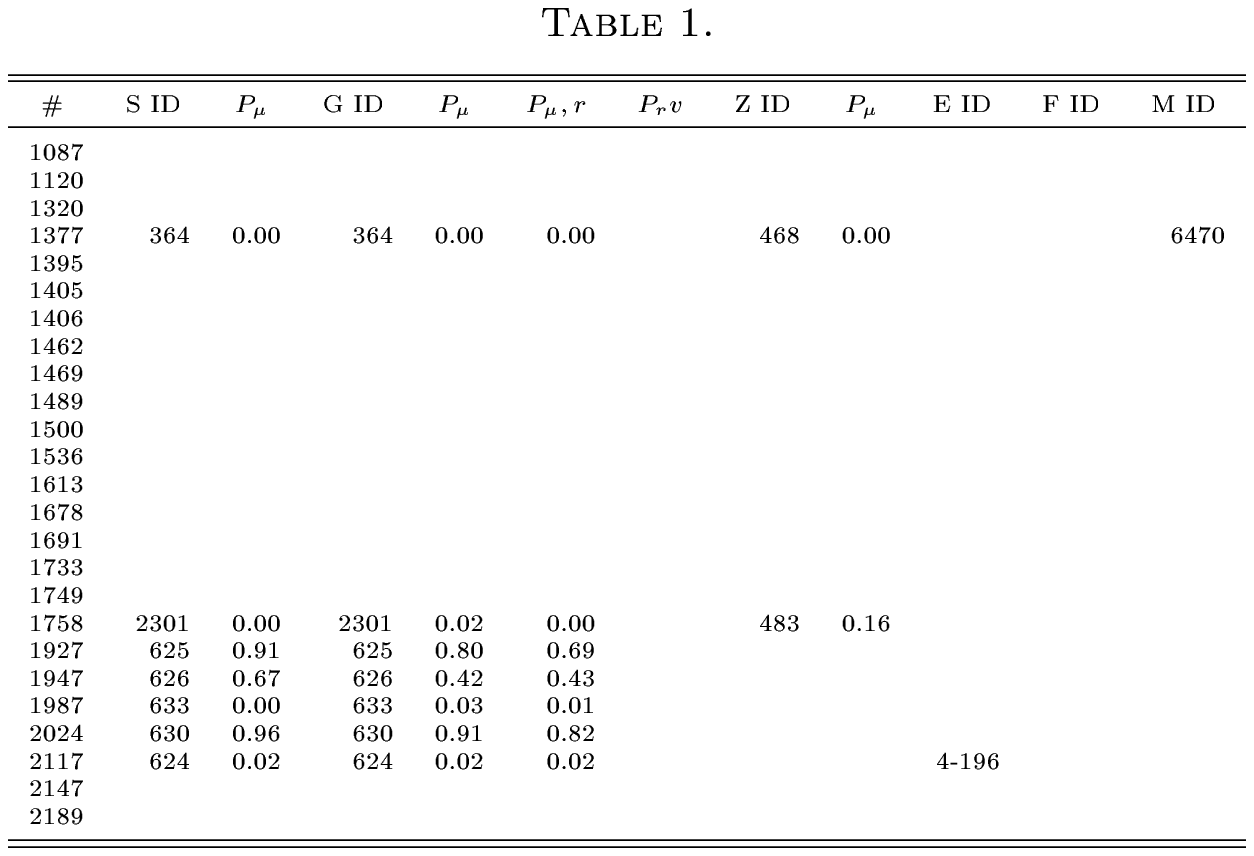}}
\caption{{\small Master list of all 977 stellar sources
included in our observations, sorted in order of increasing right
ascension, which we were able to cross-identify with the database
of \cite{fan}. From left to right: identification number (\cite{fan});
Sanders' identification number
and proper-motion cluster-membership probability (\cite{san});
Girard et al.'s identification number, proper-motion
cluster-membership probability $P_{\mu}$, proper-motion membership
probability taking into account the star's position relative to the
cluster centre $P_{\mu},{\rm r}$ and radial-velocity membership probability
$P_{rv}$ (\cite{girard}); Zhao et al. identification number and
proper-motion membership probability (\cite{zhaoea}); \cite{eggesand}
identification number; \cite{fage} identification
number; \cite{montea} identificaton number. {\it Note: Only the first 25 stars
are shown here to demonstrate the format of the table. The full table is
available only in the electronic version of the paper.}}}
\label{tab1}
\end{table*}

In Table~\ref{tab2} we list 13 stars that appear in our frames but that we
could not cross-identify in the \cite{fan} database (within
a $2\farcs5$ search radius). For these 13 stars we provide stellar 
positions derived from our data frames, as well as any cross-identifications
to other studies.

\setcounter{table}{1}
\begin{table*}[ht]
\centerline{\includegraphics[bb=35 310 610 475]{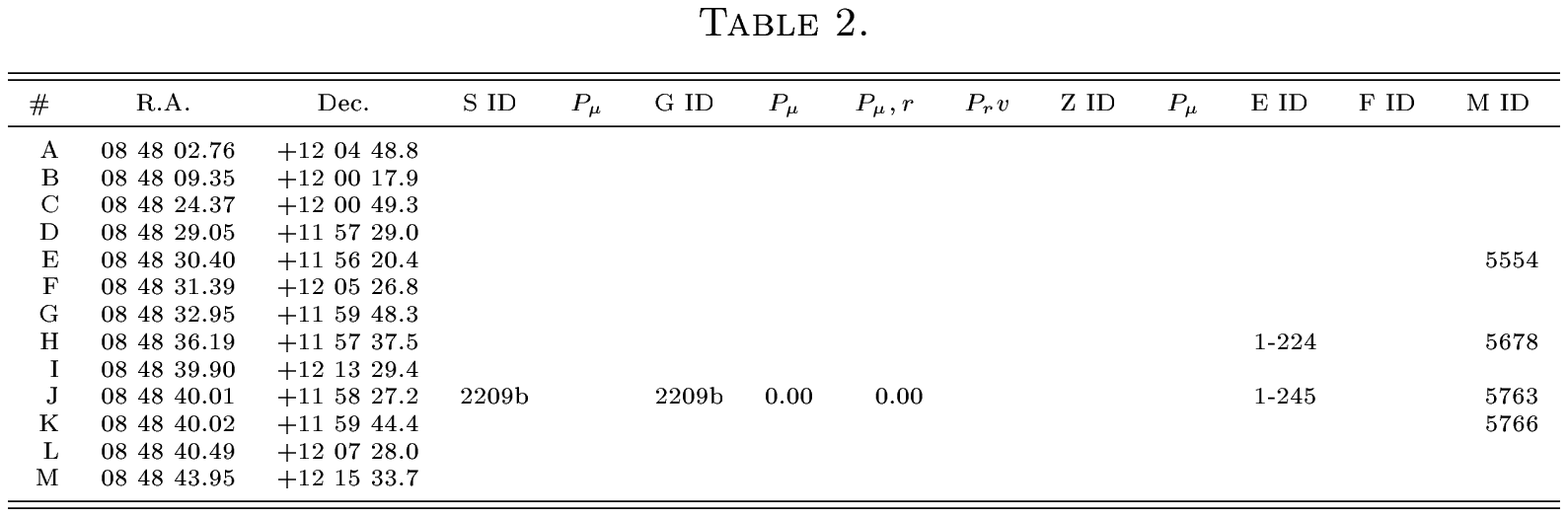}}
\caption{{\small List of 13 stellar sources included in our
observations which we were not able to cross-identify with the 
database of \cite{fan}. The stars are listed in order of
increasing right ascension. Columns are as for Table~\ref{tab1}, but also
included are right ascension and declination coordinates (equinox 1950)
as determined from our data frames.}}
\label{tab2}
\end{table*}

Given the highly inhomogeneous nature of our data (all
stars do not appear on all data frames), differential photometric
light curves are derived with the algorithm for differential
photometry of an inhomogeneous ensemble described by \cite{honn},
as described in Paper~I.

\begin{figure}[ht]
\centerline{\includegraphics[width=9.3cm]{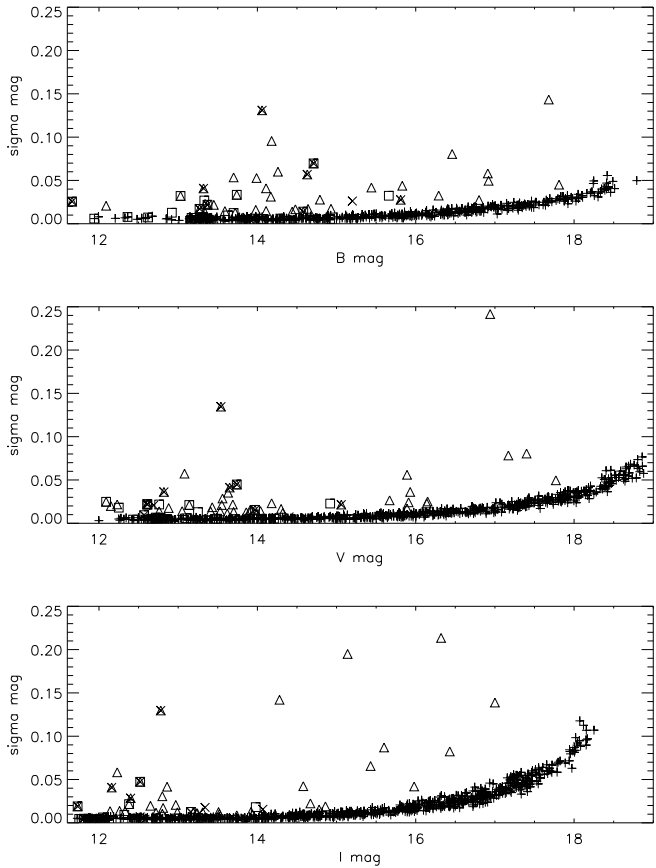}}
\caption{{\small Mean stellar magnitude versus r.m.s.\ variations in
the $B$ (top), $V$ (middle) and $I$ light curves of run 1. Variables
are indicated with triangles, spectroscopic binaries with squares,
X-ray sources with crosses. Non-variable stars are defined by the
lower envelope of points.}}
\label{m67phot2:m0sig} 
\end{figure}

The limiting precision of our differential photometry is a function of
stellar brightness and can be established from an examination of the
scatter present in the light curves of non-variable stars at each
magnitude.  The limiting photometric precision as a function of
stellar magnitude varies among the five observing runs, but generally
speaking the precision of the brightest unsaturated stars in our
exposures is flat-field limited to 5--10 mmag.  This precision level
typically holds for stars up to 2--2.5 mag fainter than the brightest
sources, and then becomes photon-noise limited and degrades for still
fainter stars. The best overall precision was achieved on our Kitt
Peak frames (run 1; see Table~1 in Paper~I).  In Fig.~\ref{m67phot2:m0sig} 
we show the r.m.s.\ variations in the $BVI$ light curves of stars from this
observing run as a function of mean stellar magnitude.  The
non-variable stars are defined by the lower envelope of points in this
figure. The brightest non-variable stars show r.m.s.\ variations of
$0.007$, $0.005$, and $0.005$ mag in $B$, $V$, and $I$, respectively.
The precision begins to degrade noticeably at around 14th mag. For the
faintest sources, at about 18.5 mag, the precision is $\sim 0.05$ mag.
We have applied a zero-point shift to the instrumental
magnitudes in each filter to roughly place our stellar magnitudes
on an absolute scale; these shifts were
determined by comparing our magnitudes to the published values of \cite{montea}. 

In Table~\ref{tab3} we present mean $BVI$ photometry for
the 990 stellar sources in our database. As discussed above,
uncertainties in the values listed are a function of stellar
brightness: formal uncertainties in the brightest sources are $0.01$
mag or less, while the faintest sources observed have formal uncertainties 
of $\sim 5$\%.  We note, however, that our photometry has not been
strictly calibrated, so that the uncertainty in the absolute photometry 
listed in Table~\ref{tab3} is more likely no better than a few percent. In
Sect.~\ref{m67phot2:results} we use these stellar magnitudes to
construct colour-magnitude diagrams for identifying objects of interest.  

\setcounter{table}{2}
\begin{table}[ht]
\centerline{\includegraphics[bb=35 250 610 535]{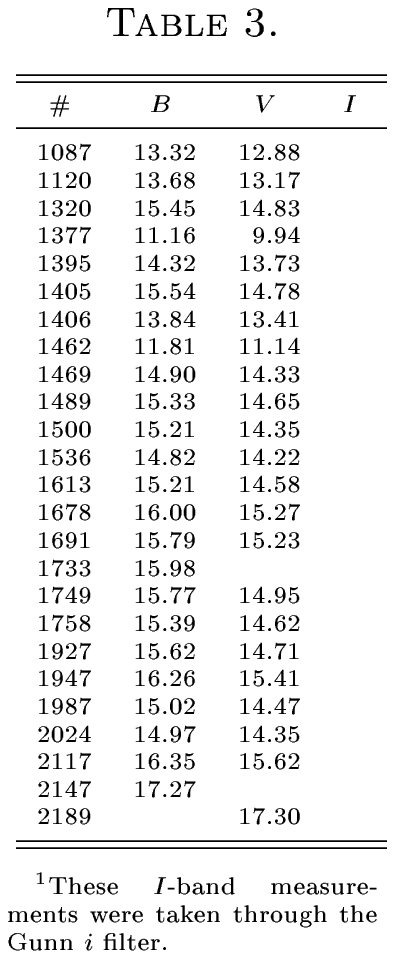}}
\caption{{\small Mean $BVI$ photometry
for the 990 stellar sources in our database. {\it Note: Only the
first 25 stars are shown to demonstrate the format of the table. 
The full table is only available in the electronic version of the paper.}}}
\label{tab3}
\end{table}

Table~\ref{tabprec} shows how our photometric precision 
varies as a function of $B-V$ and $V-I$ color for stars on the cluster main 
sequence. Note that, while these values are representative, due to the 
inhomogeneous nature of our dataset the precision achieved for any particular 
star may deviate from the values listed in the Table.

\setcounter{table}{3}
\begin{table}[th]
\centerline{\includegraphics[bb=35 320 610 476]{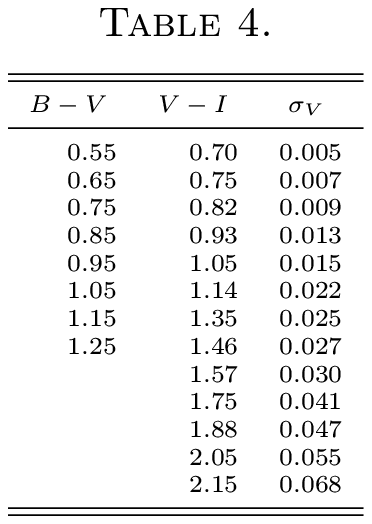}}
\caption{{\small Listing of photometric precision as a function of
main-sequence stellar colors.}}
\label{tabprec}
\end{table}

\subsection{Search for variability} \label{m67phot2:searchvar}
To identify photometric variability among the stars in our database,
we apply a $\chi^2$ test, as described in Paper I, to compute the
probability that each star's light curve is consistent with
being constant.  As our photometric precision is a function of
stellar brightness, our ability to detect low-level photometric
variations is necessarily a function of stellar brightness as well.
Among the brightest sources, the variability search is sensitive to
r.m.s.\ variations larger than $\sim 10$ mmag. Depending upon the
specific observing runs contributing data to each star's light curve,
the variability search is sensitive to variations on time scales
ranging from $\sim 0.3$ hours to $\sim 20$ days. 

Stars with data in multiple runs were analysed on a run-by-run basis,
and the results of the variability analysis for the different runs
were checked for agreement.  All instances in which the variability
analysis gives a different result in different runs can be ascribed to
differences in sensitivity between runs.

Among those stars found to be variables, we perform a Lomb-Scargle
time-series analysis (\cite{scar}) to search for the presence of a
periodic signal, following the procedure described in \cite{stass}.
For each star a periodogram is computed at 1000
frequencies between a minimum and maximum frequency corresponding to
the full time span of the light curve and one-half the typical
sampling rate, respectively. We choose the highest peak in the
periodogram as the best estimate for a possible periodicity in the
data and estimate the statistical significance of this best period by
computing a false-alarm probability (the probability that the detected
period could result from random variations). The false-alarm
probability is computed via a Monte Carlo simulation in which
periodograms are computed for 100 purely random light curves with the
same temporal sampling and photometric dispersion as the actual light curve, 
and the height of
the highest peak in these 100 test light curves is taken to correspond
to the level of 99\% significance. We report a period only if its peak
in the periodogram exceeds that of the 99\% significance level so
derived. Uncertainties in the periods are estimated as described
in Paper~I.

\section{Results} \label{m67phot2:results}

\subsection{Colour-magnitude diagrams}
\begin{figure*}[ht]
\centerline{\includegraphics[width=14.5cm]{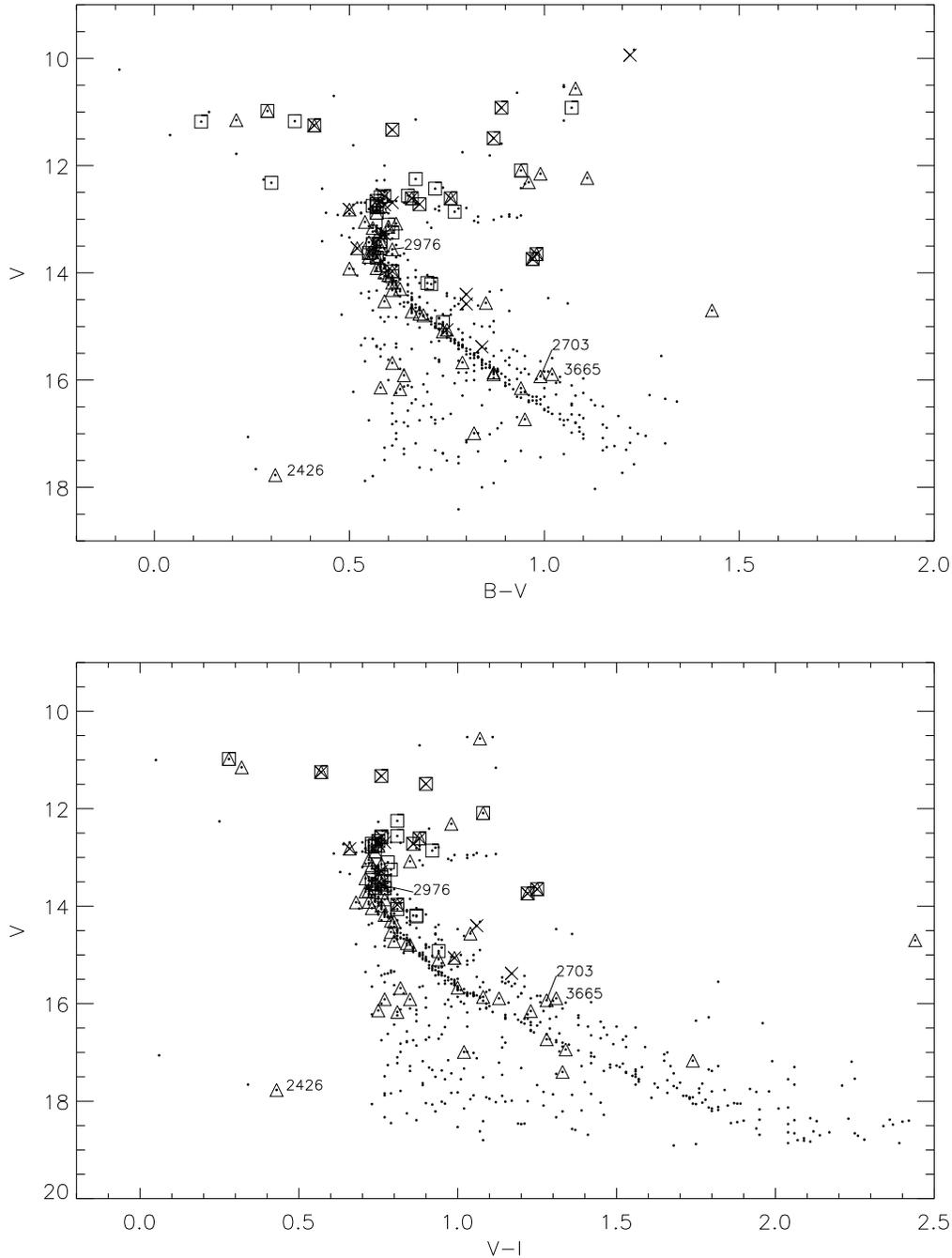}}
\caption{{\small Colour-magnitude diagrams that show $V$ versus $(B-V)$ and $V$
versus $(V-I)$ for all the stars in our observations. The variables
listed in Table \ref{m67phot2:variables} are indicated with triangles, spectroscopic binaries
with squares and X-ray sources with crosses. Periodic variables discussed
in Sect.~\ref{m67phot2:pervar} are indicated.}}
\label{m67phot2:cmd}
\end{figure*}

Though not strictly calibrated, the stellar magnitudes reported in
Table~\ref{tab3} allow us to place most of the stars in our database on a
colour-magnitude diagram. In Fig.~\ref{m67phot2:cmd} we present $V$
versus $(B-V)$ and $V$ versus \ $(V-I)$ colour-magnitude diagrams.
The colours plotted have not been de-reddened (reddening towards M\,67
is relatively small,  $E(B-V)=0.032$--0.05, \cite{nissea}, \cite{montea}).

The cluster main sequence is clearly present amid a field of apparent
non-members, extending from the cluster turnoff at $V \sim 12.5$ down
to the faint limit of our database at $V \sim 18.5$\footnote{Note that
the $B-V$ colour-magnitude diagram cuts off at $V \sim 17.5$ due to
the limiting $B$ magnitude ($B \sim 18.5$) and the stellar colours at
that magnitude ($B-V \sim 1$); see Fig.~\ref{m67phot2:cmd}.} (the range
within which our database is roughly complete).  The cluster binary
sequence is also clearly apparent along this full range.  The stars 
lying below and to the blue of the cluster main sequence have been 
noted in studies
of M\,67 before and are likely due to field stars in the halo (e.g.\
\cite{richea}).  From the observing runs intended to study
brighter sources (covering a few small areas in the cluster; see
Table~1 and Fig.~1 in Paper~I),
some blue stragglers and a portion of the giant branch 
are also present for approximately $10<V<12.5$.

\subsection{Photometric variability---General} \label{m67phot2:photvar}

In Table~\ref{m67phot2:variables} we present the 69 stars in our
photometric database that meet the criteria for photometric
variability discussed in Sect~\ref{m67phot2:searchvar}. 
Table~\ref{m67phot2:variables} also provides comments
for most of the stars listed.  These comments give additional
information such as possible periodicities, evolutionary status,
binarity, etc. Stars without comments are stars situated on 
the cluster main sequence that display only 
non-periodic variability in our observations.

Our criterion for identifying a star as variable is
that the probability of its light curve being
constant is smaller than 0.3\%. Therefore, one expects that a small
number of stars ($\sim 3$) has been classified as a variable by chance.
Statistically significant variability in more than one passband
increases our confidence that the observed variability is real. In
what follows, we refer to such stars as ``high-confidence" 
variability candidates.

To give some clue as to the nature of the stars listed in Table~\ref{m67phot2:variables}
and, ultimately, to the physical origin of
the observed photometric variability, we plot these stars in the
colour-magnitude diagrams shown in Fig.~\ref{m67phot2:cmd}.  In
addition to these variables (shown as triangles), we also indicate the
known X-ray sources (shown as X's) and the known spectroscopic
binaries (shown as squares).  The paucity of spectroscopic binaries
below $V \sim 14$ is a bias effect due to the sensitivity limit of
present spectroscopic surveys in the cluster (e.g.\ \cite{lathmathea}).
We find photometric variables in all regions of the
cluster colour-magnitude diagram; we discuss stars in each of the
various regions in more detail in Sect.~\ref{m67phot2:disc}. 

Of the 69 variables listed in Table~\ref{m67phot2:variables}, 38
are ``high-confidence" variables (i.e.\ they show variability in
more than one passband), and of these, 29 are known proper-motion members
with a probability of 75\% or greater (a total of 319 stars in our 
survey satisfy this membership criterion\footnote{We note that the
magnitude limits of the existing proper-motion surveys (i.e.\ $V \sim 16$) 
are much brighter than the faint limit of our survey ($V \sim 18.5$). 
Consequently, these variability statistics may not be representative of cluster
members fainter than $V \sim 16$.}; \cite{san}; \cite{girard}).
Of these 29 high-confidence variable members, 16 are either known
X-ray sources or binaries (or both). Another 2 stars (3780 and 4415) are
situated on the 
binary sequence. Thus, among the proper-motion members
of M\,67 surveyed by us, there remain 11/319 ($=3.4$\%) stars 
(all on the cluster main sequence) that exhibit ``high-confidence"
variability, the origin of which cannot presently be associated with
binarity and/or X-ray activity\footnote{This statement is, of course,
dependent on the sensitivity limits of the existing X-ray data. 
The ROSAT PSPC observations of \cite{belloni} have a limiting 
$L_X \approx 8 \times 10^{29}$ erg s$^{-1}$, which is roughly a factor of
$10^3$ above the X-ray luminosity of the Sun at the peak of its activity
cycle (\cite{haisch}; \cite{acton}). Thus, considerably
deeper X-ray observations will be required to definitively rule out the
presence of X-ray emission from these ``non-X-ray sources".}.

Similar numbers are derived from the observations of \cite{gillea}.
Of the 124 stars monitored by them with $V$ magnitudes within the 
limits of our study, 4 stars (not known to be X-ray sources or 
binaries) were observed to exhibit photometric
variability at levels that would have been detected by us at the
$3\sigma$ level or higher.

\subsection{Photometric variability---Periodic} \label{m67phot2:pervar}
Aside from the X-ray sources which are the subject of Paper~I, we
detected definitive periods in the light curves of only four of these
variable stars.
These are: star 2426 ($P=0.29$ days), star 2703 ($P=3.7$ days), star
2976 ($P=0.36$ days), and star 3665 ($P=0.27$ days). \\

\begin{figure*}[hbt]
\resizebox{\hsize}{!}{\includegraphics{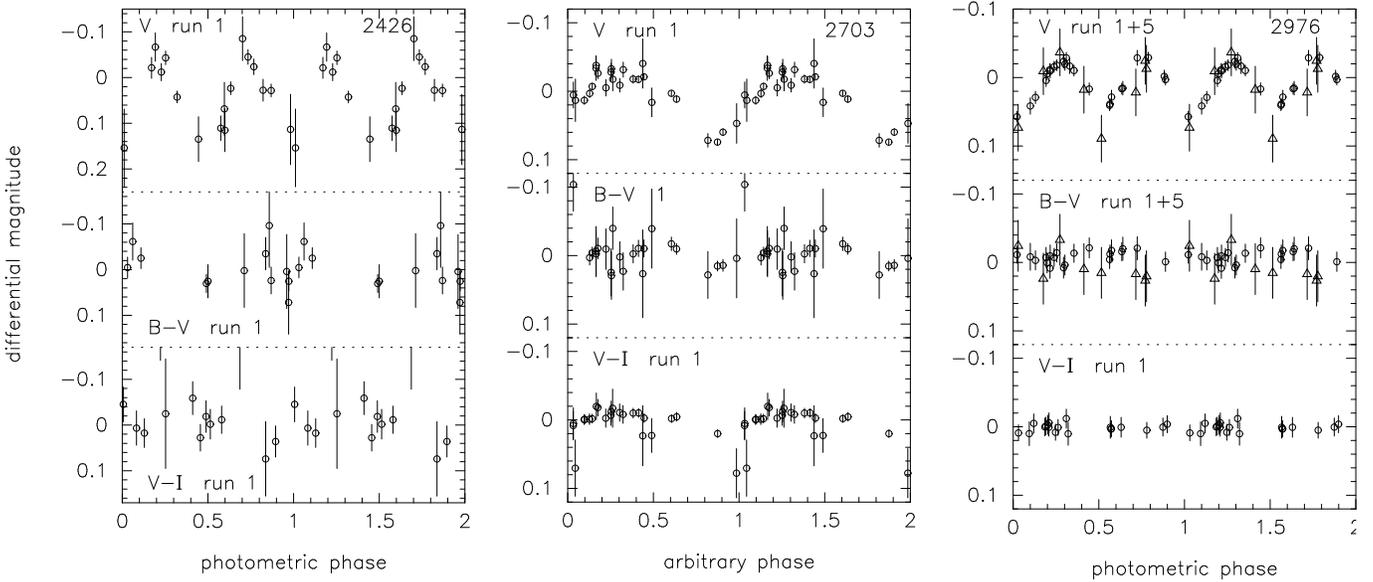}}
\caption{{\small Light and colour curves for stars 2426, 2703 and 2976
(\object{S\,757}) folded on the periods of 0.2888, 3.7 and
0.3600 days, respectively (see text, Sect.\ 3.3).  Data from different observing
runs are marked with different symbols: open circles for run 1, open
triangles for run 5.}}  \label{m67phot2:wd+ms}
\end{figure*}

\noindent
{\bf Star 2426} is located below and to the blue of the cluster main
sequence. While the best period identified by our period search (highest
peak in the periodogram) is $0.1444 \pm 0.0002$ days,
the data points at minimum light have sufficiently
large errors that it is unclear whether or not the light curve
consists of two dips with unequal depths. 
The $(B-V)$ and $(V-I)$ colour variations are not significant. 
However, additional spectroscopic and photometric observations by Orosz
et al.\ (2001, in prep.) show this star to be an Algol-type binary (with a
double-peaked light curve) with an A-star primary; the radial
velocities indicate that it is not a member of M\,67.
In Fig.~\ref{m67phot2:wd+ms} the light curve for this star is shown
folded on the period of 0.2888 days. 

\noindent
{\bf Star 2703} is located on the cluster binary sequence and exhibits
a periodic light curve with $P = 3.7 \pm 0.3$ days in $V$, and $P = 3.4 \pm 0.2$
days in $B$, both retrieved from data of run 1. The $I$ data do not show 
a statistically significant periodicity. The $V$-band light curve 
is shown in Fig.~\ref{m67phot2:wd+ms}. The $(B-V)$ and $(V-I)$ colour variations
are not significant. \\

\noindent
{\bf Star 2976} (\object{S\,757}) is located at the top of the cluster binary sequence.
The period was determined by the combined
data of runs 1 and 5 ($B$ and $V$), that together cover a time span of
2 years. The best period determined by our period search is $0.18000 \pm 0.00005$ 
days, producing a single-peaked, roughly sinusoidal light-curve.
In Fig.~\ref{m67phot2:wd+ms} the data are folded on the double
period, as we believe this star is in fact a new W\,UMa system. This 
star was first noted to be a photometric variable by
\cite{rajaea}, although no periodicity was reported. The star shows
no significant $(B-V)$ and $(V-I)$ colour variations. We discuss this
star and our reasons for labelling it a W\,UMa
system in Sect.~\ref{m67phot2:binseq}.\\

\begin{figure}[ht]
\resizebox{\hsize}{!}{\includegraphics[width=8cm,bb=275 160 535 425]{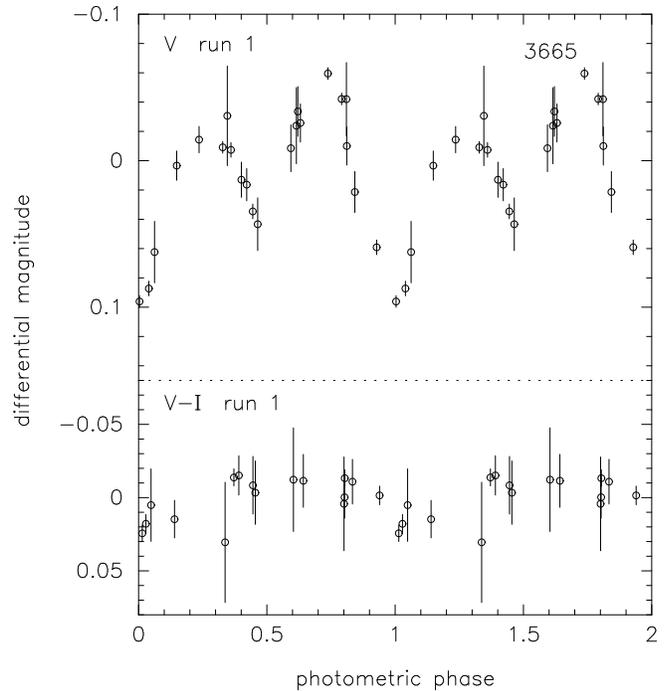}}
\caption{{\small Light and colour curves for the W\,UMa 3665
(\object{ET\,Cnc} or III-79) folded on the photometric period of 0.2712
days.}} \label{m67phot2:contact}
\end{figure}

\noindent
{\bf Star 3665} (\object{ET\,Cnc}, or III-79 in the nomenclature of \cite{eggesand})
is located on the cluster binary sequence and was 
identified as a W\,UMa system with a period of 6.49
hours by \cite{gillea}, but no error in the period was
specified. We searched for a period in a narrow window between 0.1 and
0.15 days (the power at half the period is much larger than at the
full period) and find a best period from our photometry of run 1 of
$0.1356 \pm 0.0004$ days (in $B$ and $V$, $0.1351 \pm 0.0004$ 
days in $I$) which gives a
full period of $6.51 \pm 0.02$ hours, compatible with Gilliland's measurement.
The data are folded on this period in Fig.~\ref{m67phot2:contact}.
The system is slightly redder during primary eclipse than during
secondary eclipse. This star is discussed further 
in Sect.~\ref{m67phot2:binseq}.  

\setcounter{table}{4}
\begin{table*}[ht]
\centerline{\includegraphics[width=15.5cm,bb=55 70 620 705]{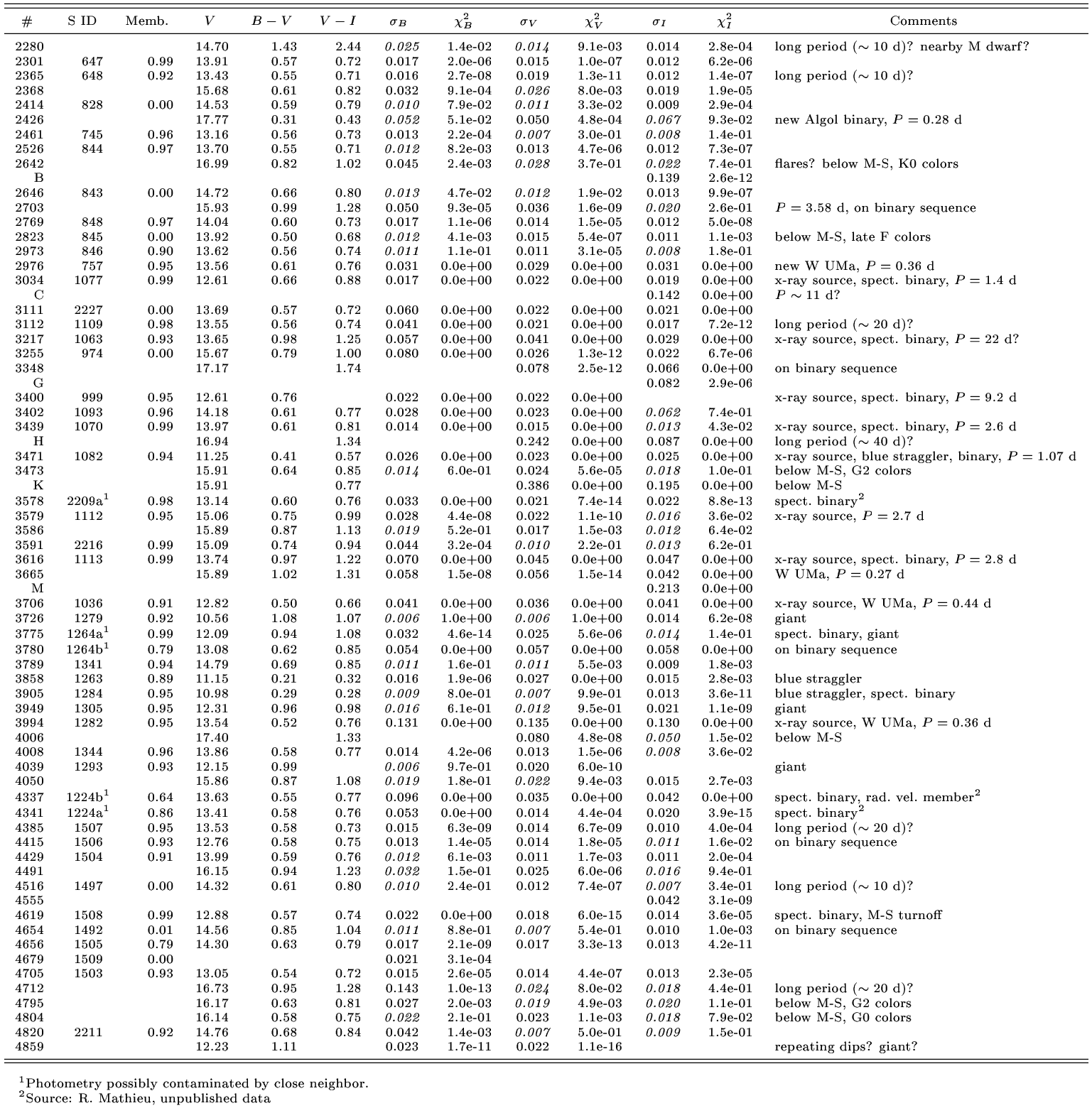}}
\caption{{\small List of variable stars. From left to right: identification
number from Table~1 and Table~2; identification number and proper-motion
membership probability from Sanders (1977) or Girard (1989) if available; average $V$
magnitude and $B-V$ and $V-I$ colours from our photometry as listed in
Table~3; r.m.s.\ of the light curves in $B$, $V$ and $I$, and corresponding
$\chi^2$ probabilities, r.m.s.\ values in
{\it italics} are not significant at the 3$\sigma$-level; comments.}}
\label{m67phot2:variables}
\end{table*}

In addition to these four periodic variables, we indicate in 
Table~\ref{m67phot2:variables} possible periods for another six stars 
(stars 2365, 3112, H, 4385, 4516, and 4712)
whose light curves show some evidence
for coherent variations on time scales longer than our observing
windows. While our period searches did not reveal statistically
significant periods for these stars, we include them here as candidate
objects for follow-up study.  We caution that these periods are
based solely on our visual impression of the light curves, and should
be taken only as a qualitative suggestion of periodicity on the timescale 
shown. Of course, these stars do still exhibit statistically
significant variability, whether or not that variability is indeed
periodic. All of these stars are on the cluster main sequence. 

We also list in Table~\ref{m67phot2:variables} tentative periods for star 2280 and star C.
Star 2280 is notable in that it is located far to
the red of the cluster main sequence in the colour-magnitude
diagram. Its colours are those of an M dwarf, so its position in the
colour-magnitude diagram suggests that it is a nearby star of the M
spectral type. Star C does not have the requisite colours to be
placed on the colour-magnitude diagram.

\section{Discussion} \label{m67phot2:disc}
We turn now to a discussion of the nature of those stars identified as
photometric variables, using the locations of these stars in the
colour-magnitude diagram as a way of organising the discussion.

\subsection{Stars below the cluster main sequence}

We have found eight variables located below the cluster main sequence.
In 7 cases, our light curves show only
irregular variations and, except for one star (star K), the amplitudes
of the variations are small (see Table~\ref{m67phot2:variables}). The available
proper-motion membership probabilities for these stars confirm 
non-membership in the
cluster (see Tables \ref{tab1} and \ref{tab2}).  
Our data provide very limited information about the nature of the observed
variability. However, at the suggestion of the referee, we have closely
examined the light curves for indications of 
RS CVn behaviour. Magnetic activity causes many RS CVn systems to be
variables, due to spots on the surface of the active star(s) that
rotate in and out of view. This usually results in periodic photometric
variability, with amplitudes ranging from hundredths to tenths of 
a magnitude, typically among stars with spectral type late-F to mid-K
(e.g.\ \cite{strass}).

We do not observe periodic variability in the light curves of
these stars. However, we cannot definitively exclude the possibility
that these are background RS CVn systems, as spots on some RS CVn can be 
variable on timescales of days or even less than a single rotation period. 
Interestingly, we note that our light curve for star 2823 (S\,845), while 
not strictly periodic, does show weak evidence for possible periodic 
behaviour (with a period of 2.9 days) for a portion of our lightcurve. 
Perhaps this is a field RS CVn with a rapidly evolving spot geometry.
We note that our light curves for two other stars (stars 4795 and 4804)
show some evidence for flaring (i.e. rapid, short-duration rises in 
brightness), but such flaring is evidently somewhat rare among RS CVn 
systems (e.g.\ \cite{henrynewsom}). 

While no proper-motion membership information is available for the one
periodic variable below the main sequence (star 2426, see Sect.~\ref{m67phot2:pervar}), 
the observations of Orosz et al.\ (2001, in prep.) indicate that this 
star is an Algol-type binary and a non-member.

\subsection{Stars on the cluster main sequence}
In Sect.~\ref{m67phot2:pervar} we mentioned six stars situated on
the cluster main sequence that show some evidence for periodic
variability.  If these tentative periods are upheld, they may be
related to the stellar rotation periods (due to, e.g., starspot
modulation).  In addition, we have found another 21 stars on the 
cluster main sequence (not including the X-ray sources discussed
in Paper~I) whose light curves
show non-periodic variations at a statistically significant level in
at least one filter (Table~\ref{m67phot2:variables}). 
The amplitudes of variability exhibited by a few of these stars is
remarkably large (few tenths of a mag; see Fig.~\ref{m67phot2:m0sig}).
Our data do not shed much light on the nature of
this variability. Perhaps the variations we observe have their origin
in magnetic activity on the surfaces of these G and K stars.

Fifteen of these 27 stars are high-confidence variables, 
showing statistically significant variability in more than one filter. 
Of these, 11 are known proper-motion members of the cluster (see
Sect.~\ref{m67phot2:photvar}). As discussed in Sect.~\ref{m67phot2:photvar}, 
photometric variability at the levels to which our survey is sensitive 
($\sim$ 1--2\%) is evidently rare among the main-sequence members of M\,67,
with an occurrence rate of at most a few percent in those observed by us.

\subsection{Stars on the cluster binary sequence} \label{m67phot2:binseq}
Photometric variables on the cluster binary sequence are interesting
because such variability can point the way to the discovery of
interacting binary systems.  If periodic, the observed variability may
be related to the dynamics of the binary system.  Three of
the variables for which we found a photometric period are situated on
the cluster binary sequence.

The two most notable variable stars on the cluster binary
sequence in our observations are the two W\,UMa (contact binary)
systems, stars 3665 and 2976, which have already been introduced in 
Sect.~\ref{m67phot2:results}. The former (\object{ET Cnc}) was discovered 
by \cite{gillea} while the latter is newly discovered here.
\object{ET\,Cnc} has a primary-eclipse depth in $V$ of 0.16 mag, and a 
secondary-eclipse depth of 0.1 mag.
The unequal eclipses in the observed
light curve potentially indicate that the system is in poor thermal
contact or is semi-detached (see also discussion in Paper~I on the
X-ray source W UMa \object{S\,1036}).

We have discovered that star 2976 (\object{S\,757}) is a strong candidate for
being a W\,UMa system in M\,67.  \cite{san} gives this star a
proper-motion membership probability of 95\%. 
The most likely period from our time-series
analysis is 0.1800 days, which we interpret as the half-period of
0.3600 days, assuming two eclipses of
very similar depth.  Spectroscopic radial-velocity measurements will
be needed to confirm this period. Nonetheless, a period of 0.36 days
places this star on the W\,UMa period-colour relation (e.g.\ \cite{rucibook})
very well, given its $(B-V)$ colour of 0.61.
We note that \cite{shetrone} have also recently reported detection of 
eclipses in this star with a period of $P \sim 0.4$ days.

The discovery of this new W\,UMa variable brings the total number of
such contact binaries in M\,67 to four. We can estimate the frequency
of W\,UMa systems in M\,67 by comparing this number to the number of
proper-motion cluster members from, e.g., the study of \cite{girard}.
The Girard et al.\ study includes 367 proper-motion members
(probability $\ge 75$\%) among stars with $V<15.5$ in a region 
$34' \times 42'$ about the cluster centre. 
This yields a W\,UMa frequency of $4/367 = 1.1$\%, which is consistent
with the contact-binary frequency in other Galactic open clusters of $\sim$ 1\% 
(\cite{rucinski}). We note that while the new 
W\,UMa discovered by us is contained in the Girard et al.\ study, one
of the three previously known systems is not (\object{ET Cnc} is a bit too
faint with $V=15.8$), even though it is within the spatial boundaries
of that study. And the spatial area of the present study is somewhat
smaller than that of the Girard et al.\ study. As noted by \cite{rucinski},
a detailed accounting of contact-binary statistics is made difficult 
because of these differences in depth and spatial coverage of different
studies. Even so, the W\,UMa frequency in M\,67 is evidently of order 1\%.

In addition to these two contact binary systems, we have also
discovered periodic variability in another star on the cluster binary
sequence, star 2703 with a period of 3.6 days. While our data permit us
to say little about this star, its location in the colour-magnitude
diagram and its periodic light curve make this a prime candidate for
follow-up spectroscopic monitoring.  Perhaps the observed photometric
period corresponds to the binary orbital period. No membership
information is presently available for this star.

Four stars (3348, 3780, 4415, 4654) situated on the
cluster binary sequence did not evince periodic variability in
our data, but nonetheless showed statistically significant variability.
Star 4415 (\object{S\,1506}) is a proper-motion member and was
monitored by Mathieu et al.\ (1986) for radial-velocity variations, 
but no indication of binarity was found ($\sigma=1$ km s$^{-1}$ in 8
observations spanning 2 years); perhaps this is a wide binary.
Star 4654 (\object{S\,1492}) is a proper-motion non-member, and in any case
its photometric variability is only significant in one filter.
The other 2 stars, both of which are ``high-confidence" variables, 
deserve spectroscopic follow-up to determine if
they are interacting binary systems. Star 3780 (\object{S\,1264b}) is a
proper-motion member, but we note that its photometry might be
affected by the presence of a close neighbor. Star 3348
does not presently possess cluster membership information.

\begin{figure}[ht]
\resizebox{\hsize}{!}{\includegraphics[width=8cm,bb=62 404 550 710]{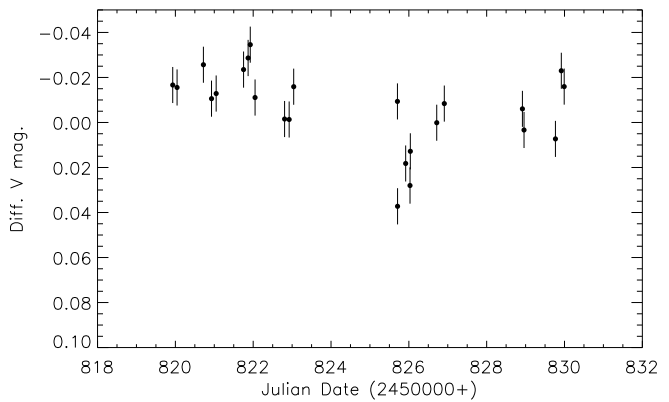}}
\caption{{\small $V$-band light curve for the spectroscopic binary, star 4619
(\object{S\,1508}).}}
\label{m67phot2:nm}
\end{figure}

Finally, we have found four stars (3578, 4337, 4341, 4619)
showing statistically significant 
variability that are also known spectroscopic binaries (R.\ Mathieu,
private communication). 
Star 4337 (\object{S\,1224b}) has an orbital period of 12.44 days and an
eccentricity of 0.03. Star 4341 (\object{S\,1224a}) has an orbital period of
726 days, and an eccentricity of 0.3.  
For star 3578 (\object{S\,2209a}) no orbital solution has yet been derived. 
The light curves of
these three stars show no evidence for periodicity on the orbital
or pseudo-synchronous periods (Hut 1981). However, we caution that our
photometry for these stars may be contaminated by close neighbors.
Star 4619 (\object{S\,1508}),
situated at the cluster turnoff, was found by \cite{mathlathea}
to be a spectroscopic binary with an orbital period of 25.9 days and
an eccentricity of 0.44.  The $BVI$ light curves of this star do not
show evidence for periodic variability on the spectroscopic orbital
period or the pseudo-synchronous period of 11.2 days. This star's 
$V$-band light curve is shown in Fig.~\ref{m67phot2:nm}.

\subsection{Giant stars}
Five of the variables reside on the red giant branch of the cluster
colour-magnitude diagram.
These variables
show relatively low levels of variability, with r.m.s.\ $\sim
0.02$ mag. We note that for three of these five stars, the observed
variability is statistically significant in only one of the filters
observed (i.e.\ they are not ``high-confidence" variables). 
For the other giant stars included in our observations, we can
place 3$\sigma$ upper limits on variability of $\sim 0.015$ mag, based
on the limiting photometric precision of our photometry for the
brightest sources.

Star 3775 (\object{S\,1264a}) is a spectroscopic binary, and should be monitored
further to study any possible connection between the binary orbit and
photometric variability. While a ``high-confidence" variable, we caution 
that our photometry is suspect due to the presence of a close neighbor.
\cite{mathea86} monitored
the giants 3726 (\object{S\,1279}), 4039 (\object{S\,1293}), and 3949 
(\object{S\,1305}), but found no
evidence for radial-velocity variations ($\sigma \leq$ 0.5 km s$^{-1}$
in about 15 observations spanning more than 10 years, for all three
stars).  Thus we consider it unlikely that the variability, if real, 
is related to binarity.  \cite{henrea} found a large fraction of
low-amplitude ($\sim 0.01$ mag) photometric variables among a sample
of 187 G ($\sim 25$\%) and K ($\sim 50$\%) giants. For the group of
giants of type earlier than K2---which includes our four variable
giants---they exclude both radial pulsation and rotational
spot-modulation as the origin of the brightness variations. Henry et
al.\ suggest that non-radial, $g$-mode pulsations most likely give rise
to the variability.

\begin{figure}[t]
\resizebox{\hsize}{!}{\includegraphics[width=8cm,bb=62 404 550 710]{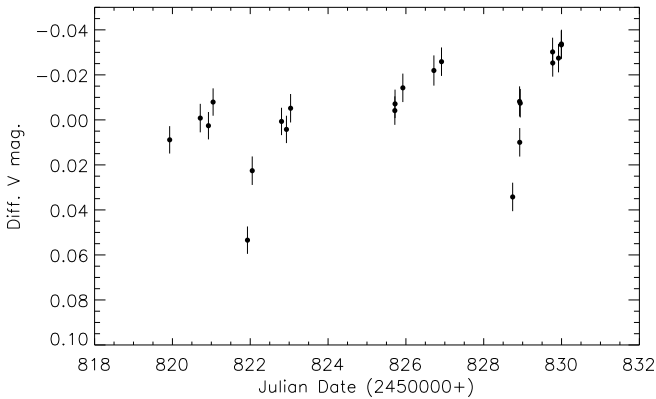}}
\caption{{\small $V$-band light curve for the giant star 4859.}}
\label{m67phot2:giant}
\end{figure}

Interestingly, star 4859, the other high-confidence variable on the
giant branch, shows a peculiar light curve, with rapid,
short-duration dips (Fig.~\ref{m67phot2:giant}). Unfortunately, 
being situated in the far outskirts of the cluster, no
membership information is available for this star.

\subsection{Blue stragglers}
Two of the blue stragglers in M\,67 not detected in X-rays show
statistically significant photometric variability in our
data. Photometric variability among blue stragglers is particularly
interesting in light of the uncertain evolutionary status of these
objects. Photometric variability may be a clue to the presence of a
binary companion (e.g. in the case of the eclipsing blue straggler
\object{S\,1082}, see \cite{goraea}; \cite{vdbergea2001a})
or may provide information on the stellar
structure of the stars (in the case of oscillations, see discussion in
Gilliland \&\ Brown 1992).  


Star 3905 (\object{EX\,Cnc} or \object{S\,1284}) is a spectroscopic binary that shows
low-amplitude photometric variations with a period of $\sim 1.3$ hours
first discovered by \cite{gillea}, \cite{gillbrow}, and \cite{simo}.
The $B$ and $V$ light curves we obtained during the highly
sampled run 4 show a similar behaviour. These short-timescale
variations are probably related to the star's position within the
$\delta$ Scuti instability strip and not to the orbital period of 4.2
days (\cite{milolath92a}).


Star 3858 (\object{S\,1263}) was monitored spectroscopically by \cite{milolath92b},
but no orbit was determined. Gilliland et al.\ included
this star in their search for solar-analog oscillations, but 
observed an r.m.s.\ scatter of only 0.005 mag.
\cite{simo} similarly found no
evidence for photometric variability, but \cite{kimea} do report
a high dispersion in the light curve of this star. The light curves
that result from our observations also display a large scatter (up to
0.03 mag), and this variability is statistically significant in all 
three bands. While our results and those of Simoda and Kim et al.
can be explained in terms of sensitivity differences of the 
studies, the Gilliland et al.\ result suggests that the variability
observed by us and by Kim et al.\ does not persist at all times.

\section{Conclusions} \label{m67phot2:summ}

Our survey of photometric variability among 990 stars in the old open
cluster M\,67 detected 69 variable stars.  
Among the brightest sources in our sample, detection of variability 
$\sigma \approx 10$ mmag (with $> 3\sigma$ confidence) is achieved;
for the typical star observed, sensitivity to variability at levels
$\sigma \approx 20$ mmag is achieved.
Membership information is
available for 439 stars (46 variables) included in our observations
and marks 319 (38 variables) as members with a probability of at least
75\% (\cite{san}; \cite{girard}). Of these 38 variable cluster
members, 29
exhibit variability in more than one of the passbands used, increasing
our confidence that the observed variability is real. Nine of these
stars are periodic variables.

In all cases the amplitude of variability is low, ranging from a few
hundredths to a few tenths of a magnitude. Our study is sensitive to
brightness variations on time scales of 0.3 hours to $\sim 20$
days. Apparently, at the age of M\,67 variability on these
time scales and at these amplitudes is strongly associated with
binarity, as 14 of the 29 ``high-confidence" variable members are known binaries. 
One of the other high-confidence variables (\object{S\,1112}) is an X-ray source
for which binarity has not been established, another is a
blue straggler (\object{S\,1263}), and still two others are situated on the 
cluster binary sequence.
{\it Periodic} variability is especially rare for single stars, 
for in 8 of the 9 periodic variable members observed by us (including 
the new candidate contact system \object{S\,757}),
7 of which are X-ray sources, the variability finds its origin in the
binary nature of the stars (eclipses, ellipsoidal variations,
rotational spot-modulation in tidally locked binaries). This confirms
the picture that rapid rotation in an old population can only be
maintained in close binaries.

In the contact binary 3665 (\object{ET\,Cnc}), for which no membership
information exists, the variability is the result of binarity as well.
We encourage spectroscopic observations of the three remaining stars
that exhibit periodic variations (the faint and blue star 2426 and star
2703 on the binary sequence, both without membership information) and
the member 3579 (\object{S\,1112}, discussed in paper I) to establish their
binary status and/or obtain an indication for membership from their
radial velocity.  

Also, more observations should be obtained of the
stars for which we provide tentative periods, in the first place to
further examine if their photometric variability is indeed periodic
and secondly to establish if they are single or binary.

The origin of the photometric variability for the remaining stars
discussed in this paper is in most cases unknown. As possible causes
for the variations we suggest low-level surface activity, stellar
pulsations or, especially for the stars on the binary sequence, binary
interaction.

\begin{acknowledgements}
The authors wish to thank Magiel
Janson, Rien Dijkstra, Gertie Geertsema, Remon Cornelisse and Gijs
Nelemans for obtaining part of the data used in the paper.  The Kitt
Peak National Observatory is part of the National Optical Astronomy
Observatories, which is operated by the Association of Universities
for Research in Astronomy, Inc. (AURA) under cooperative agreement
with the National Science Foundation. The Jacobus Kapteyn Telescope is
operated on the island of La Palma by the Isaac Newton Group in the
Spanish Observatorio del Roque de los Muchachos of the Instituto de
Astrofisica de Canarias. The Dutch 0.91-m Telescope was operated at La
Silla by the European Southern Observatory. Support for this work was
provided by NASA through Hubble Fellowship grant HST-HF-01144.01-A
by the Space Telescope Science Institute, which is operated by AURA,
for NASA, under contract NAS 5-26555.
This research made extensive use of the WEBDA Open Cluster Database
developed and maintained by J.-C. Mermilliod.  MvdB is supported by
the Netherlands Organization for Scientific Research (NWO).
\end{acknowledgements}

\end{document}